%% file: a0_int_submit.tex
 \documentclass[twocolumn,10pt,prl,preprint,floatfix,nofootinbib,superscriptaddress,showpacs,amssymb]{revtex4}
 \usepackage{graphicx}
 \def\deg{^\circ}
 \input{prd_definitions}
\begin{document}

%\pagewiselinenumbers
%\doublespace

 \bibliographystyle{apsrev}

%%%%%%%%%%%%%%%%%%%%%%%%%%%%%%%%%%%%%%%%%%%%%%%%%%%%%
 \title {Phenomenological interpretation of the  multi-muon
         events reported by the CDF collaboration }
%%%%%%%%%%%%%%%%%%%%%%%%%%%%%%%%%%%%%%%%%%%%%%%%%%%%%
 \affiliation{Laboratori Nazionali di Frascati, Istituto Nazionale 
              di Fisica Nucleare, I-00044 Frascati, Italy\\}
 \affiliation{University of Cyprus, Nicosia CY-1678, Cyprus\\}
 \affiliation{Duke University, Durham, North Carolina  27708\\}
 \affiliation{University of Illinois, Urbana, Illinois 61801\\}
 \author{P.~Giromini}
 \affiliation{Laboratori Nazionali di Frascati, Istituto Nazionale
              di Fisica Nucleare, I-00044 Frascati, Italy\\}
 \author{F.~Happacher}
 \affiliation{Laboratori Nazionali di Frascati, Istituto Nazionale
              di Fisica Nucleare,  I-00044 Frascati, Italy\\}
 \author{M.~J.~Kim}
 \affiliation{Laboratori Nazionali di Frascati, Istituto Nazionale
              di Fisica Nucleare,  I-00044 Frascati, Italy\\}
 \author{M.~Kruse}
 \affiliation{Duke University, Durham, North Carolina  27708\\}
 \author{K.~Pitts}
 \affiliation{University of Illinois, Urbana, Illinois 61801\\}
 \author{F.~Ptohos}
 \affiliation{University of Cyprus, Nicosia CY-1678, Cyprus\\}
 \author{S.~Torre}
 \affiliation{Laboratori Nazionali di Frascati, Istituto Nazionale
              di Fisica Nucleare, I-00044 Frascati, Italy\\}
%%%%%%%%%%%%%%%%%%%%
% \linenumbers
 \begin{abstract}
 We present a phenomenological conjecture of new physics that is suggested
 by the topology and kinematic properties of the multi-muon events recently
 reported by the CDF collaboration. We show that the salient features of 
 the data can be accounted for by postulating the pair production of
 three new states $h_1$, $h_2$, and $h_3$ with masses in the range
 of 15, 7.3, and 3.6 $\gevcc$, respectively. The heavier states 
 cascade-decay into the lighter ones, whereas the lightest state 
 decays into a $\tau$ pair with a lifetime of the order of 20 ps.  
 \end{abstract} 
%%%%%%%%%%%%%%%%%%%%
 \preprint{FERMILAB-PUB-08-231-E}
 \maketitle
 The Higgs mechanism provides a scheme for the electroweak symmetry breaking
 (EWSB). Electroweak precision tests (EWPT) indicate that the standard model
 (SM) Higgs boson is light, $m_h < 186 \; \gevcc$~\cite{ewgroup}, with a 
 central value considerably below the lower bound of $114\; \gevcc$ from 
 direct searches~\cite{hlim}. While it is possible that one is misled in
 interpreting the EWPT data in terms of a light Higgs mass, the EWSB sector
 in the SM Lagrangian appears to be the most elusive and the one most likely
 to provide experimental surprises~\cite{gunion,fullana,gunion1,hidval}.
 In Refs.~\cite{bbxs,a0disc}, the CDF collaboration presents a set of studies
 of multi-muon events.
  Reference~\cite{bbxs} uses multi-muon events to measure the correlated $b\bar{b}$
 production cross section. That study is extended in Ref.~\cite{a0disc} to additional
 properties of multi-muon events.  Based on the
 properties of these events, this paper explores a possible conjecture
 of new physics that might
 relate the results of  Ref.~\cite{a0disc} to the EWSB mechanism.

 The data set used in Refs.~\cite{bbxs,a0disc} has been acquired with a 
 dedicated dimuon trigger, and consists of events containing two central
 ($|\eta|<0.7$) muons, each with transverse momentum $p_T \geq 3 \; \gevc$,
 and with invariant mass larger than 5 $\gevcc$. References~\cite{bbxs,a0disc}
 show that, when both trigger (initial) muons arise from particles
 that have decayed within the beam pipe of radius 1.5 cm, their number and
 kinematic properties  are
 correctly predicted by a SM simulation. According to the simulation, 
 approximately 96\% of the known sources of dimuons, such as Drell-Yan,
 $\Upsilon$, $Z^0$, and heavy flavor production, satisfy this condition.
 The sum of these contributions (1131090 events) is 
 for simplicity referred to as QCD production~\cite{a0disc}.
 Reference~\cite{a0disc} also reports the observation of
 295481 events, referred to as ghost events,
 in which at least one muon originates beyond the beam pipe.
 A large fraction of these events 
 is attributed to muons arising from $\pi$, $K$, $K^0_S$, and hyperon
 decays as well as to secondary interactions in the detector volume. 
 However, a small but significant fraction of these events has characteristics
 that cannot be explained by known processes in conjunction with the
 current understanding of the CDF~II detector, trigger, and event
 reconstruction. Reference~\cite{a0disc} has investigated the rate and kinematic
 properties of tracks and muons contained in a $36.8^{\deg}$ cone around the
 direction of each trigger muon in ghost events.
 The nature of these events is characterized
 by four main features. The impact parameter~\cite{ip} distribution of 
 initial muon pairs is markedly different from that of QCD events.
 In $36.8^{\deg}$ ($\cos \theta \geq 0.8$) cones around the initial muon direction,
 the rate of additional muons and charged tracks is significantly higher than that
 of QCD events. The invariant mass of the initial and additional muons
 distributes differently from that expected from sequential semileptonic decays
 of hadrons with heavy flavor. The distributions of the
 impact parameter of additional muons, and of the distance
 between the $p\bar{p}$ collision point and
 secondary vertices reconstructed using  pairs of tracks
 contained in a cone,
 do not correspond to the lifetime of any known particle.

 Although it is not excluded that the features of these events can be
 later explained in terms of known sources~\cite{a0disc},
 this paper presents a phenomenological conjecture of new 
 physics that shows that it is possible to account for the various features
 of these events, and also suggests additional observations that are further
 tested with the published data~\cite{a0disc}. In this conjecture, a fraction of the ghost
 events is attributed to pair production of three new states $h_1$, $h_2$,
 and $h_3$ with masses in the range of 15, 7.3, and 3.6 $\gevcc$,
 respectively. The heavier states cascade-decay into the lighter ones,
 whereas the lightest state decays into a $\tau$ pair with a lifetime
 of the order of 20 ps~\cite{a0disc}.

\begin{figure}
 \begin{center}
 \vspace{-0.2in}
 \leavevmode
 \includegraphics*[width=0.5\textwidth]{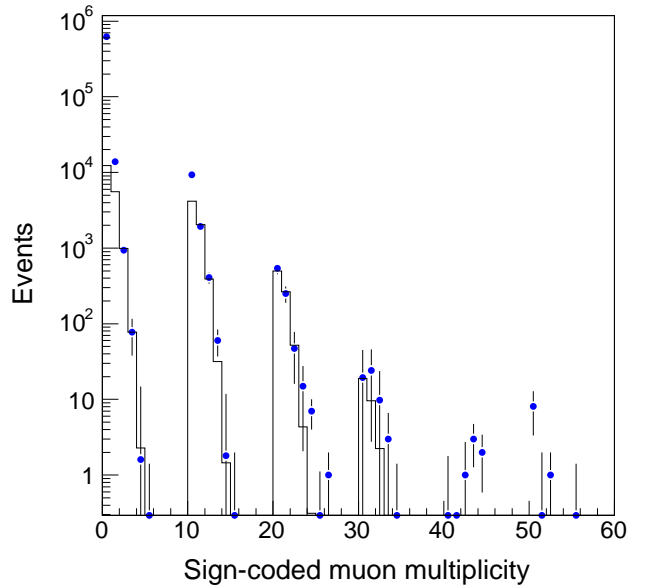}
 \caption[]{Sign-coded multiplicity distribution of additional muons found
            in a $\cos \theta \geq 0.8$ cone around the direction of a
            primary muon in ghost events. The points are reproduced from
            Ref.~\cite{a0disc}, in which the QCD contribution has been
	    removed and the distribution has been corrected for the fake muon
            contribution. An additional muon increases the multiplicity by
            1 when it has opposite and by 10 when it has same sign charge as
            the initial muon. The first bin represents cones without additional muons.
         As examples, the third bin indicates cones
            with 3 muons with charge ($+-\;-$) or ($-++$); and the 21st bin
            indicates cones with 3 muons with charge ($+++$) or ($-\;-\;-$).
	    The solid line is the prediction of the toy-simulation of a decay
            into eight $\tau$ leptons (see text).}
 \label{fig:fig_14}
 \end{center}
 \end{figure}

 Figure~\ref{fig:fig_14}, reproduced from Ref.~\cite{a0disc}, shows the
 sign-coded multiplicity distribution of additional muons contained in a
 $36.8^{\deg}$ cone around the direction of an initial muon in ghost 
 events. In the plot, an additional muon increases the multiplicity by 1
 when of opposite sign and by 10 when of the same charge of the initial muon.
 Leaving aside the case in which no additional muons are found, an increase
 of one unit in the muon multiplicity corresponds in average to a population
 decrease of approximately a factor of seven. This factor happens to coincide
 with the inverse of the $\tau \rightarrow \mu$ branching fraction (0.174)
 multiplied by the 83\% efficiency of the muon detector, and suggests
 that these muons might arise from $\tau$ decays with
 a kinematic acceptance close to unity. We use Monte Carlo pseudoexperiments
 to model the shape of the muon multiplicity distribution in 
 Fig.~\ref{fig:fig_14}. The pseudoexperiments generate 4 $\tau^-$ + 4 $\tau^+$
 leptons and decay them into muons with a 17.4\% probability. In the
 pseudoexperiments, initial and additional muons are identified with the
 measured~\cite{a0disc} detector efficiencies of 50\% and 83\%, respectively.
 The pseudoexperiment result is shown in Fig.~\ref{fig:fig_14}, in which the
 simulated distribution is normalized to the integral of the data for
 multiplicity bins higher than 10. The comparison of the muon multiplicity
 distribution in ghost events with the toy simulation suggests that
 approximately 13200 events contain 8 $\tau$ leptons inside a
 $\cos \theta \geq 0.8$ cone. One interpretation is that they are
 8-$\tau$ decays of objects $h_1$ that are relatively light with 
 respect to the transverse momentum with which they are produced.
 %%%%%%%%%%%%%%%%%%%%%%%%%%
%%%%%%%%%%%%%%%%%%%%%%%%% 

\begin{figure}
 \begin{center}
 \vspace{-0.3in}
 \leavevmode
 \includegraphics*[width=0.5\textwidth]{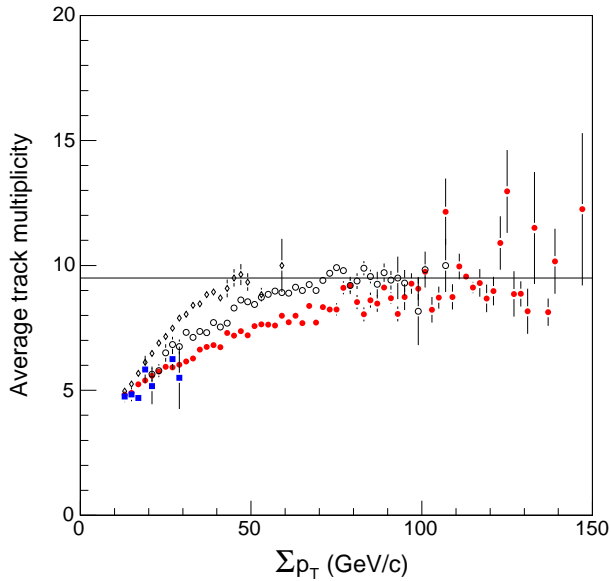}
 \caption[]{Average number of tracks in a $36.8^{\deg}$ cone around the
            direction of a primary muon as a function of $\sum p_T$, the
            transverse momentum of all tracks. Data ($\bullet$), reproduced
            from Ref.~\cite{a0disc}, are compared to a 
            $H \rightarrow h_1 h_1$ simulation with
            $m_{h_1}=15$, and ($\diamond$) $m_H=115$ or ($\circ$) 
            $m_H=300 \; \gevcc$. The transverse momentum distribution of
            the data is different from that of the simulations, but the average
            number of tracks has the same asymptotic value, as indicated by
            the straight line. The QCD prediction ({\tiny $\blacksquare$}) is
            based on the few events predicted by the heavy flavor simulation,
            normalized to the number of initial muon pairs in the data and
            implemented with the probability that tracks mimic a muon signal.}
 \label{fig:fig_17}
 \end{center}
 \end{figure}

 Next, we test the data for an obvious consequence of the
 $h_1 \rightarrow 8 \tau$ conjecture. If the hypothesis is correct, one 
 expects that, when  the $h_1$ transverse momentum is very large, the $8\tau$
 decays produce an average of 9.5 tracks with $p_T \geq 2 \; \gevc$ in a 
 $36.8^{\deg}$ cone. We compare the data presented in Ref.~\cite{a0disc}
 to simulated events which contain $h_1$ states produced with large 
 transverse momenta. We use the {\sc pythia} Monte Carlo program~\cite{pythia} 
 to generate fictitious $ p \bar{p} \rightarrow H \rightarrow h_1 h_1$ events
 followed by $h_1 \rightarrow 8 \tau$ decays~\cite{sim}. We use this process
 for convenience, without implying that the $h_1$ states might be Higgs
 bosons or might arise from Higgs decays. Figure~\ref{fig:fig_17} compares
 the average track multiplicity in the data and in some of the simulations.
 In order to reject the QCD contribution, Reference~\cite{a0disc} considers 
 events that contain at least three muons in a $36.8^{\deg}$ cone. The figure
 shows the average number of tracks with $p_T\geq 2 \; \gevc$ contained in
 a 36.8$^{\deg}$ cone around a primary muon as a function of the total
 transverse momentum of the tracks. The asymptotic value of the average
 track multiplicity in the data adds support to the conjecture of a
 $h_1 \rightarrow 8 \tau$ decay initially suggested by the muon multiplicity
 distribution. 
%%%%%%%%%%%%%%%%%%%%%%%%%%
%%%%%%%%%%%%%%%%%%%%%%%%% 

\begin{figure}
 \begin{center}
 \vspace{-0.3in}
 \leavevmode
 \includegraphics*[width=0.5\textwidth]{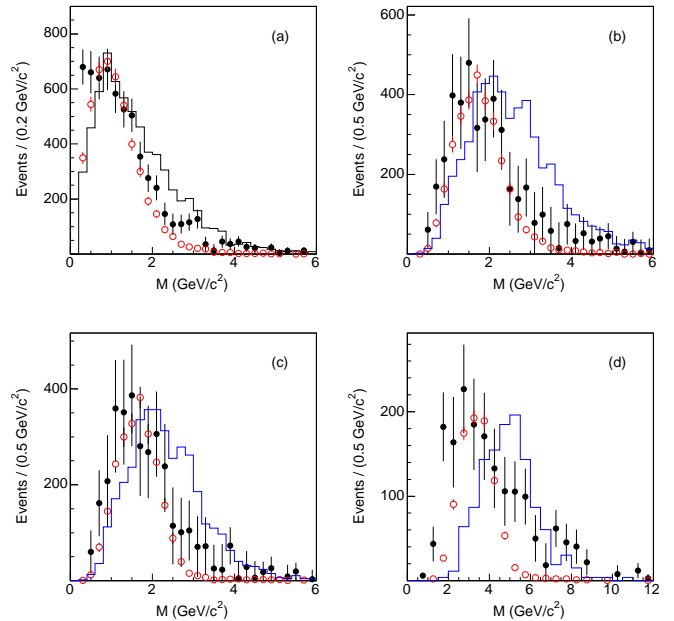}
 \caption[]{Invariant mass, $M$, distributions of all muons in a $36.8^{\deg}$
            cone when (a) both cones contain at least two muons, (b) a cone
            contains three or more muons, (c) a cone contains three muons,
            and (d) of muons and tracks for cones containing 5 to 6 tracks
            and three or more muons. The fake muon contribution has been
            subtracted. The invariant mass distributions predicted by
            a simulation with $m_{h1}=15$ ($\circ$) and $m_{h1}=20 \; \gevcc$
            (histogram) are superimposed to the data ($\bullet$) reproduced
            from Ref.~\cite{a0disc}.  } 
 \label{fig:fig_18}
 \end{center}
 \end{figure}

 In the assumption that leptons and tracks contained in a $36.8^0$ cone
 around a primary muon arise from $h_1 \rightarrow 8 \tau$ decays, the
 mass of the $h_1$ particle is determined by comparing data and simulations
 in the following invariants proportional to $m_{h_1}$: (a) the invariant
 mass of all muons in a cone when both cones contain at least two muons;
 (b) the invariant mass of all muons in a cone containing at least three  
 muons; (c) the muon invariant mass for cones containing exactly three muons;
 (d) the invariant mass of muons and tracks for cones containing three or
 more muons and 5 to 6 tracks. Reference~\cite{a0disc} shows that the request
 of a large number of muons suppresses the QCD contribution. In cases when
 fewer muons are requested, the QCD background is larger and has been 
 subtracted in Ref.~\cite{a0disc}. These invariant mass distributions are
 shown in Fig.~\ref{fig:fig_18}. The data are compared to simulations of
 the process $H \rightarrow h_1 h_1$ with $m_H = 300 \; \gevcc$, and
 $m_{h_1} =15$ and $20 \; \gevcc$, respectively. A mass of 15 $\gevcc$,
 close to the invariant mass of 8 $\tau$ leptons, provides a fair modeling
 of the data, whereas a mass of 20 $\gevcc$ appears to be too high. 
%%%%%%%%%%%%%%%%%%%%%%%%%%

\begin{figure}
 \begin{center}
 \vspace{-0.3in}
 \leavevmode
 \includegraphics*[width=0.5\textwidth]{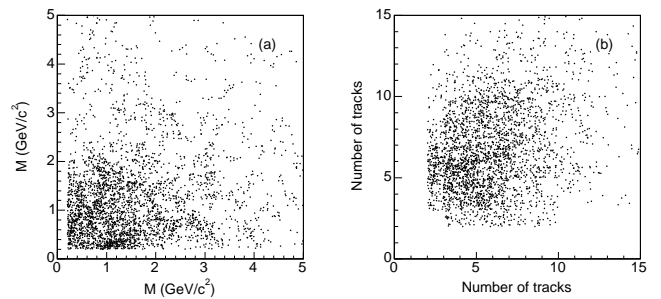}
 \caption[]{Two-dimensional distributions, reproduced from Ref.~\cite{a0disc},
            of (a) the invariant mass, $M$, of all muons and (b) the total
            number of tracks contained in a $36.8^{\deg}$ cone when both
            cones contain at least two muons.}
 \label{fig:fig_19}
 \end{center}
 \end{figure}

\begin{figure}
 \begin{center}
 \vspace{-0.3in}
 \leavevmode
 \includegraphics*[width=0.5\textwidth]{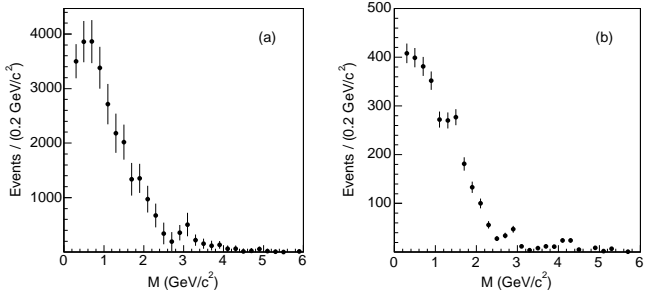}
 \caption[]{ Distributions, reproduced from Ref.~\cite{a0disc}, of the
             invariant mass, $M$, of all muons contained in (a) the 27990
             $36.8^{\deg}$ cones with two or more muons and (b) the 3016
             events in which both cones contain two or more muons. QCD and
             fake muon contributions have been removed.}
 \label{fig:fig_20}
 \end{center}
 \end{figure}

 We investigate if the data are consistent with the hypothesis of $h_1$ pair
 production by studying the rate and properties of events in which two
 $36.8^{\deg} $ cones contain a muon multiplicity larger than that of
 QCD events. In the sample of ghost events isolated by the study
 in Ref.~\cite{a0disc}, there are $27990\pm761$ cones containing two or
 more muons, $4133 \pm 263$ cones containing three or more muons, and
 $3016 \pm 60$ events in which both cones contain two or more muons.
 Figure~\ref{fig:fig_19} plots two-dimensional distributions of the
 invariant mass of all muons and of the number of tracks contained in
 each cone for the 3016 events.
%%%%%%%%%%%%%%%%%%%%%%%%%%
%%%%%%%%%%%%%%%%%%%%%%%%% 
 Figure~\ref{fig:fig_20} shows that the invariant mass distribution
 of all muons contained in the 27990 cones containing at least two
 muons is consistent with that of the 3016 events in which both
 cones contain at least two muons~\cite{a0disc}.  
%%%%%%%%%%%%%%%%%%%%%%%%%%
%%%%%%%%%%%%%%%%%%%%%%%%% 
 We compare the data to a simulation of the process 
 $f\bar{f} \rightarrow h_1 h_1$ generated with the {\sc pythia} Monte Carlo
 program, in which the $h_1$ pair production is mediated by a photon exchange.
 In 8549600 generated events, we find 15580 events with a pair of initial
 muons that pass the analysis selection of Ref.~\cite{a0disc}. These simulated events contain 
 7997 cones with two or more muons, and there are 1044 events in which both 
 cones contain two or more muons. The ratio 1044/7997 is in quite good 
 agreement with that of the data (3016/27990)~\cite{sim1}. Based on these comparisons,
 a fraction of the ghost events seems to be more consistent with the
 conjecture that $h_1$ states are pair produced.  

\begin{figure}
 \begin{center}
 \vspace{-0.3in}
 \leavevmode
 \includegraphics*[width=0.25\textwidth]{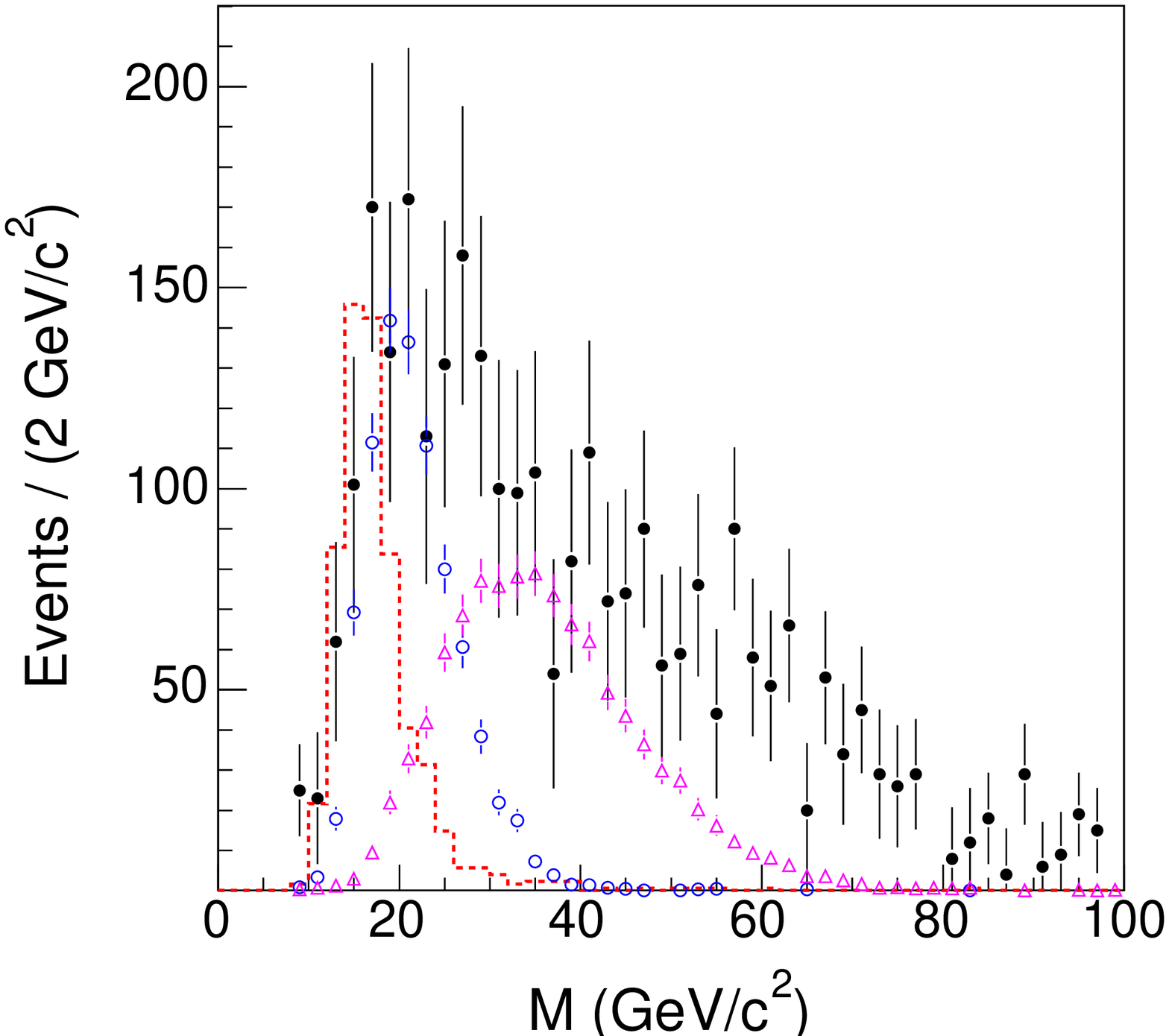}\includegraphics*[width=0.25\textwidth]{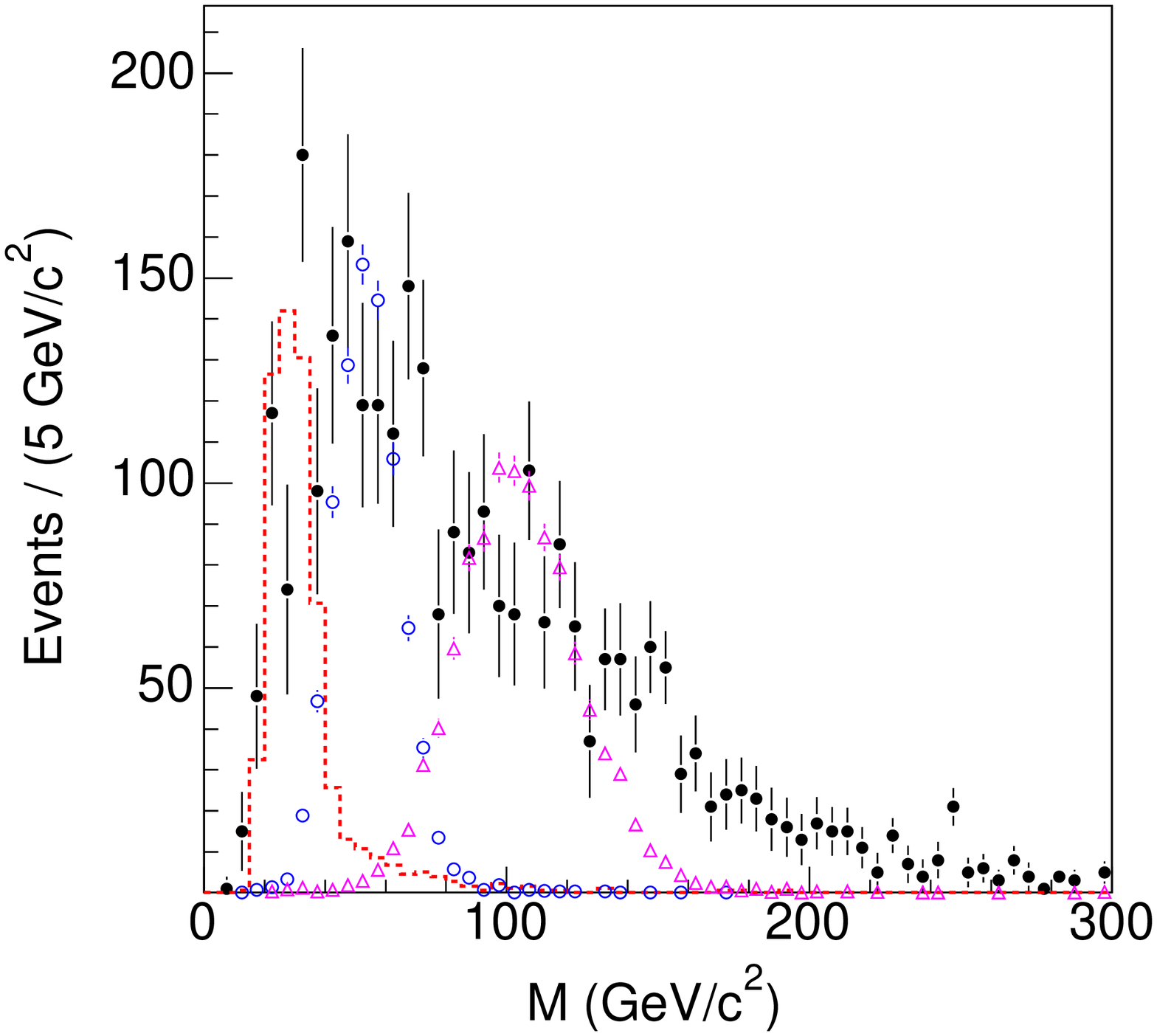}
 \caption[]{Invariant mass, $M$, distribution of (left) all muons and (right)
            all tracks with $p_T \geq 2 \; \gevc$ for events in which both 
            cones contain at least two muons. QCD and fake muon contributions
            have been removed. Simulations of the processes
            $f\bar{f} \rightarrow h_1 h_1$ (dashed histogram), and
            $H \rightarrow h_1 h_1$ with $m_H=150$ ($\circ$) and $300\; \gevcc$
            ($\triangle$) are superimposed to the data ($\bullet$) reproduced
            from Ref.~\cite{a0disc}.}
 \label{fig:fig_22}
 \end{center}
 \end{figure}

 The dependence of the $h_1$ pair production on the Mandelstam variable
 $\hat{s}$ is studied by comparing data to different  simulations in the
 following distributions shown in Fig.~\ref{fig:fig_22}: (a) the invariant
 mass of all muons for events in which both cones contain at least two muons;
 and (b) the invariant mass of all tracks for events in which both cones
 contain at least two muons.
%%%%%%%%%%%%%%%%%%%%%%%%%%
%%%%%%%%%%%%%%%%%%%%%%%%% 
 The data distributions do not show any evidence of resonant production.
 However, the data distributions fall less rapidly than in the
 $f\bar{f} \rightarrow h_1 h_1$ simulation, a possible indication that the 
 $h_1$ pair production is not mediated by a photon or a gluon.
 The $f\bar{f} \rightarrow h_1 h_1$ distribution shown in Fig.~\ref{fig:fig_22}
 corresponds to a production cross section of 3.4 nb. Events in which the
 $h_1$ pairs have larger invariant mass are modeled with a simulation of 
 the process $p\bar{p} \rightarrow H \rightarrow h_1 h_1$ with $m_H=150$
 and $300 \; \gevcc$. The results of these simulations, also shown in
 Fig.~\ref{fig:fig_22}, correspond to a production cross section of 50 and
 35 pb for $m_H=150$ and 300 $\gevcc$, respectively.
%%%%%%%%%%%%%%%%%%%%%%%%%%%%%
 Using the acceptances measured with these different simulations we again
 estimate that approximately 5\% of the 295481 ghost events with two initial
 muons can be explained by pair production of $h_1$ particles.

In the assumption that ghost muons are produced in the decays of these states,
 the fits to the high impact parameter tail of additional muons in ghost
 events~\cite{a0disc}, where the distribution is well modeled by an exponential function,
 can be used to estimate  their lifetime to be $21.4 \pm 0.5$ ps. If the $h_1$ states decay
 directly into 8 $\tau$ leptons with a lifetime of 20 ps, the impact
 parameters of muons due to ghost events would be highly correlated since
 the $\tau$ lifetime is negligible compared to that of the $h_1$ states.
 In contrast, the study in Ref.~\cite{a0disc} finds a very weak correlation
 between the impact parameters of muons contained in the same cone.
 We compare data from Ref.~\cite{a0disc} to simulations in order to test
 the more elegant conjecture of a three-stage decay
 $h_1 \rightarrow 2\;h_2 \rightarrow 4\;h_3 $, where $h_3$ is the
  particle decaying into $\tau$ pairs. 
 We use  simulated samples  of the processes $f\bar{f} \rightarrow h_1 h_1$
 and $p\bar{p}\rightarrow H\rightarrow h_1 h_1$ with $m_H=300 \; \gevcc$,
 in which the $h_1$ states decay into 8 $\tau$
 leptons through a three-stage decay.
  We attribute in turn a 20 ps
 lifetime to only one of the $h_1$, $h_2$, and $h_3$ states. We measure the
 correlation between the impact parameters of muons contained in a 
 $36.8^{\deg}$ cone for the different cases. The correlation factor is
 $\rho_{d_p d_s}=0.39 $, 0.15, and 0.05 when the lifetime is attributed to
 the $h_1$, $h_2$, and $h_3$ states, respectively, whereas it is measured to
 be 0.03 in ghost events~\cite{a0disc}.
 This indirect  method provides evidence for the possible
 existence of two additional states, $h_2$ and $h_3$. The latter state is
 long-lived and decays into $\tau$ pairs. Following the assignment of a
 15 $\gevcc$ mass to the $h_1$ state, one expects the $h_2$ mass to be in
 the range $7.1-7.5$ $\gevcc$ and the $h_3$ mass $\simeq 3.6 \; \gevcc$. 
 The observed number and properties of the ghost events can  accommodate
 the additional pair production of at least one of the $h_2$ and $ h_3$ states.
 When the 20 ps lifetime is
 attributed to the $h_3$ state, only a few percent of the simulated events 
 selected as in Ref.~\cite{a0disc} survive the additional
 request that both initial muons originate inside the beam pipe. This models
 closely what happens in the data, as reported in
 Ref.~\cite{a0disc}. 

 Because the production mechanism of these hypothetical states is not
 understood, we cannot use the simulation to improve the measurement of
 the $h_3$ lifetime reported in Ref.~\cite{a0disc}. However, we 
 highlight one difficulty that arises if ghost events are due to mixed
 pair production of $h_1$, $h_2$, and $h_3$ states. In Ref.~\cite{a0disc},
 the lifetime of the muon parent particle has been estimated by fitting
 with an exponential function
 the muon impact parameter distributions in the range $0.5-2.0$ cm. 
 We have produced simulated samples of $h_n$ pair production using the processes
 $f\bar{f} \rightarrow h_n h_n$ and
 $p\bar{p} \rightarrow H \rightarrow h_n h_n$ with $m_H=300 \; \gevcc$ 
 and $n=1,2,3$. In these simulations, states heavier than the $h_3$ particles
 cascade-decay into them and ultimately produce  events with 4, 8, and
 16 $\tau$ leptons in the final state. We generated several samples with 
 $h_3$ lifetimes ranging from 10 to 40 ps. For simulated events due to $h_3$
 or $h_2$ pair production, analogous fits to the impact parameter distributions
  return the lifetime value used
 in the event generation. However, in the simulation of $h_1$ pair production,
 some of the $h_1$ cascade-decay products are not relativistic and the decay
 kinematics has greater complexity. As a consequence, the  fits to 
 the impact parameter distributions
 underestimate the lifetime by approximately 30\%.

\begin{figure}
 \begin{center}
 \vspace{-0.3in}
 \leavevmode
 \includegraphics*[width=0.5\textwidth]{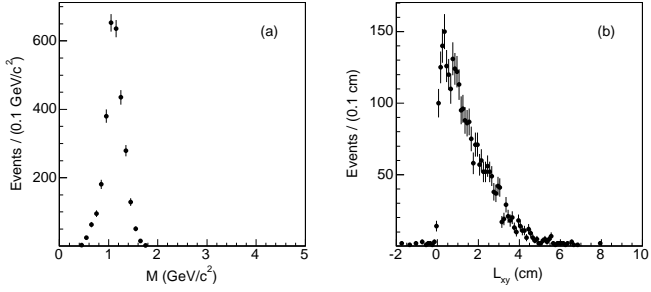}
 \caption[]{Distributions of the (a) invariant mass and (b) distance $L_{xy}$
            (see text) of three-track systems in events simulated with the 
            process $p\bar{p} \rightarrow H \rightarrow h_1 h_1$ with 
            $m_H=150 \; \gevcc$. The three-track systems are produced by
            single $\tau$ decays into three hadrons. } 
 \label{fig:fig_32}
 \end{center}
 \end{figure}

 The identification of $\tau$ decays into three hadrons would provide
 additional support to the conjecture of new physics used to account for
 a fraction of the ghost events. This decay channel could also be used 
 for a complementary measurement of the $h_3$ lifetime by using the
 $L_{xy}$ distribution of the three-track secondary vertices, where
 $L_{xy}$ is the distance between the secondary
 vertex  and the $p\bar{p}$ collision
 point projected onto the transverse momentum of the three-track
 system. Reference~\cite{a0disc} has searched the data for such secondary
 vertices, and has verified the detector response by using identified
 $K_S^0 \rightarrow \pi^+ \pi^- $ decays.
 That study~\cite{a0disc} reconstructs  
 three-prong secondary vertices by using tracks with $p_T \geq 1.0 \; \gevc$ and
 $|\eta| \leq 1.1$ in a $36.8^{\deg}$ cone around the direction of 
 each initial muon. Track systems with total charge of $\pm 1$ are 
 constrained to arise from a common space point. 
 In this study, we use similar criteria to search for
 hadronic $\tau$ decays in simulated events, and compare our result to the
 published data~\cite{a0disc}. In a sample  generated 
 with the process $p\bar{p} \rightarrow H \rightarrow h_1 h_1$ with
 $m_H=150 \; \gevcc$, there are in average two $\tau$ decays
 into three hadrons per event. Approximately 8\% of these decays, or 0.16 $\tau$
 hadronic decays per event, survive these selection criteria.
 Figure~\ref{fig:fig_32} shows the invariant mass and $L_{xy}$ distributions
 of the three-track systems that are also identified at generator level as
 $\tau$ decays into three hadrons.
 %%%%%%%%%%%%%%%%%%%%%%%%%%
%%%%%%%%%%%%%%%%%%%%%%%%% 
 
\begin{figure}
 \begin{center}
 \vspace{-0.3in}
 \leavevmode
 \includegraphics*[width=0.5\textwidth]{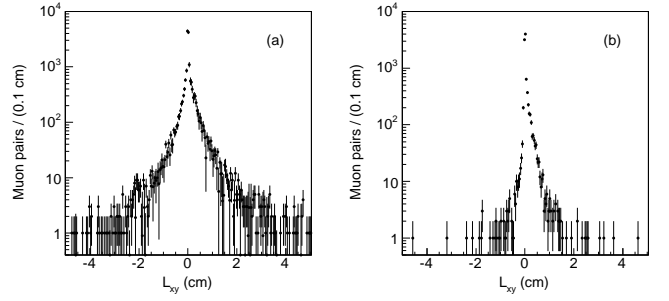}
 \caption[]{Distribution, reproduced from Ref.~\cite{a0disc}, of the distance
            $L_{xy}$ of fit-constrained vertices of three-track systems
            contained in a $36.8^{\deg}$ cone around the direction of an 
	    initial muon for (a) ghost and (b) QCD events. We select cases
            in which angular cones contain only three tracks.}
 \label{fig:fig_37}
 \end{center}
 \end{figure}

\begin{figure}
 \begin{center}
 \vspace{-0.3in}
 \leavevmode
 \includegraphics*[width=0.5\textwidth]{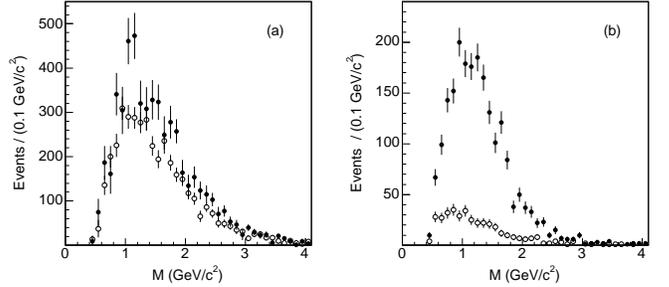}
 \caption[]{Distributions,  reproduced from Ref.~\cite{a0disc},
            of the invariant mass, $M$, of three-track systems
            in (a) ghost and (b) QCD events. Systems with distance 
            $L_{xy} \geq 0.04$ cm ($\bullet$) are compared to those with 
            $L_{xy} \leq -0.04$ cm ($\circ$). Ghost events with positive 
            $L_{xy}$ exhibit an excess of events with the expected shape
            of  $\tau$ into three hadron decays.}
 \label{fig:fig_38}
 \end{center}
 \end{figure}

 The simulation also contains 5.5 three-track combinations per event that
 pass the same selection criteria, and the signal of the 3-hadron $\tau$
 decays is swamped by the combinatorial background.
 However, this signal is comparable to the combinatorial
 background in a simulated sample of $h_3$ pair
 production in which a 36.8$^{\deg}$ cone contains sometimes
 one muon and three tracks from the two $\tau$ decays.
 In this case, the vertices of the three-track systems identify
 correctly the $h_3$ decay vertex and generate an excess of events at
 positive  $L_{xy}$ distances.
 Reference~\cite{a0disc} presents a subsample of ghost events
 in which 
 a  36.8$^{\deg}$ cone around the
 direction of an initial muon contains only three tracks with
 $p_T \geq 1 \; \gevc$. Three-track systems with total charge of $\pm 1$
 are constrained to arise from a common space point.
 Figure~\ref{fig:fig_37} shows the $L_{xy}$ 
 distribution for ghost and QCD events \cite{a0disc}. Ghost events show an
 excess at positive $L_{xy}$ that is, however, not as large as that of QCD
 events in which most of the muon plus the three-track combinations arise
 from single $b$-quark decays. Figure~\ref{fig:fig_38} compares the invariant
 mass of the three-track systems for positive and negative $L_{xy}$
 values~\cite{a0disc}. In the case of ghost events, the invariant mass of
 three-track systems with positive $L_{xy}$ exhibits an excess of events
 consistent with the  shape of $\tau$ decays into three hadrons shown in
 Fig.~\ref{fig:fig_32}.
%%%%%%%%%%%%%%%%%%%%%%%%%%
%%%%%%%%%%%%%%%%%%%%%%%%%%
%%%%%%%%%%%%%%%%%%%%%%%%%%
%%%%%%%%%%%%%%%%%%%%%%%%

 In conclusion, we suggest one possible phenomenological interpretation
 of the  multi-muon events recently
 reported by the CDF collaboration. As shown by the comparisons between
 the published data and simulations based on our conjecture, the most interesting 
 features of these events can be accounted for by postulating the pair
 production of three new states $h_1$, $h_2$, and $h_3$ with masses in
 the range of 15, 7.3, and 3.6 $\gevcc$, respectively. The heavier states 
 cascade-decay into the lighter ones, whereas the lightest state decays
 into a $\tau$ pair with a lifetime of the order of 20 ps. The mechanism
 that produces $h_1$ pairs is completely obscure. It does not appear to
 be resonant nor mediated by a photon or gluon exchange.
 The observed pair production cross section ($\simeq 100$ nb) is a 
 few orders of magnitude larger
 than what is predicted if the $h_n$ states belonged 
 to the Higgs sector~\cite{gunion,hidval}.

 We thank the Fermilab staff and the CDF collaboration institutions for
 their contributions. This work was supported by the U.S. Department
 of Energy and National Science Foundation and the Italian Istituto
 Nazionale di Fisica Nucleare. We are indebted to S.~Mrenna for guiding
 us through several subtleties of the {\sc pythia} Monte Carlo program.
%%%%%%%%%%%%%%%%%%%%%%%%%%%%%
%%%%%%%%%%%%%%%%%%%%%%%%%%%%%%%

\input{bibliography_int_prl.tex}
%%%%%%%%%%%%%%%%%%%%%%%%%%%%%%%
%%%%%%%%%%%%%%%%%%%%%%%%%%%%%%%
 \end{document}

%% file: prd_definitions.tex
\def\deg{^\circ}

\def\Z0{${\em Z^0\/}$}

\def\r#1 {$^{#1}$}

\newcommand{\gevc} { {\rm GeV/}c}
\newcommand{\gevcc}{ {\rm GeV/}c^2}

\def\gepsfcentered#1{
  \def\testit{#1}
  \def\lbracket{[}
  \ifx\testit\lbracket
    \let\dofilecmd=\gepsfwithopt
  \else
    \let\dofilecmd=\gepsfnoopt
  \fi
  \dofilecmd}

\def\gepsfnoopt#1{
  \begin{center}
  \leavevmode
  \epsffile{#1}
  \end{center}}

\def\gepsfwithopt#1 #2 #3 #4]#5{
  \begin{center}
  \leavevmode
  \gepsfmaxx=0.94\textwidth
  \epsffile[#1 #2 #3 #4]{#5}
  \end{center}}

\newdimen\gepsfmaxx
\def\epsfsize#1#2{
  \ifnum \epsfxsize=0
    \ifnum \epsfysize=0
      \ifnum #1 > \gepsfmaxx
        \gepsfmaxx
      \else
        #1
      \fi
    \else
      \epsfxsize
    \fi
  \else
    \epsfxsize
  \fi
}

%% file: a0_int_submit.bbl
\begin{thebibliography}{99}
\bibitem{ip}      The impact parameter is the distance of closest approach of
                  a track to the primary event vertex in the plane transverse 
                    to the beam direction.
\bibitem{sim}     We have generated several simulated samples with a Higgs
                  mass, $m_H$, ranging from 115 to 300 $\gevcc$ using the 
                  option {\sc isub=151} of the version 6.4 of the {\sc pythia}
                  Monte Carlo program. We have also generated pair production
                  of Higgs bosons ($f\bar{f} \rightarrow h_n h_n$) with the
                  option {\sc isub=300}. The $h_n$ widths are set to be negligible
                  with respect to the detector resolution. 
                   We use the interface {\sc pyslha.f}
                  and decay Higgs bosons according to phase space. The $\tau$
                  decays are simulated with the {\sc tauola}
                  program~\cite{tauola}
\bibitem{sim1}    In a different simulation, that uses the process
                  $p\bar{p} \rightarrow H \rightarrow h_1 h_1$ with 
                  $m_H =300 \; \gevcc$, this ratio is a factor of two higher
                  than in the data because heavy resonant states are decayed
                  by the {\sc pythia} generator program into $h_1$ pairs with
                  well balanced transverse momenta.
\end{thebibliography}
